%% file: main.tex
\documentclass{article}

\usepackage{spconf,amsmath,graphicx,hyperref}
\usepackage{amssymb,amsfonts}
\usepackage{algorithmic}
\usepackage{url}
\usepackage{multirow}
\usepackage{booktabs}
\usepackage{textcomp}
\usepackage{epstopdf}
\usepackage{xcolor}
\usepackage{float}
\usepackage{placeins}
\usepackage{orcidlink}
\usepackage{makecell}
\usepackage{etoolbox}

\let\oldsection\section
\renewcommand{\section}[1]{%
    \vspace{-8pt}%
    \oldsection{#1}%
    \vspace{-7pt}%
}

\let\oldsubsection\subsection
\renewcommand{\subsection}[1]{%
    \vspace{-5pt}%
    \oldsubsection{#1}%
    \vspace{-5pt}%
}

\let\oldsubsubsection\subsubsection
\renewcommand{\subsubsection}[1]{%
    \vspace{-4pt}%
    \oldsubsubsection{#1}%
    \vspace{-3pt}%
}

\apptocmd{\thebibliography}{%
    \setlength{\itemsep}{0pt}%
    \setlength{\parsep}{0pt}%
    \setlength{\parskip}{0pt}%
}{}{}

\title{
CRAF: Cross-View Residual-Aware Fusion for Deepfake Speech Detection
}

\name{
Minh-Xuan Phan\orcidlink{0009-0006-8462-6538},
Khalid Zaman\orcidlink{0009-0004-0809-7537},
Candy Olivia Mawalim\orcidlink{0000-0001-9853-8893},
Masashi Unoki\orcidlink{0000-0002-6605-2052}
}

\address{
Japan Advanced Institute of Science and Technology
}

\begin{document}

\maketitle


\setlength{\abovedisplayskip}{2pt plus 1pt minus 1pt}
\setlength{\belowdisplayskip}{2pt plus 1pt minus 1pt}
\setlength{\abovedisplayshortskip}{1pt plus 1pt}
\setlength{\belowdisplayshortskip}{1pt plus 1pt minus 1pt}
\setlength{\jot}{2pt}

\input{sections/abstract}

\input{sections/introduction_v2}

\input{sections/data}

\input{sections/methodology_v4}

\input{sections/experiments}

\input{sections/conclusion}


\newpage

\bibliographystyle{IEEEbib}
\bibliography{strings,refs}

\end{document}

%% file: sections/abstract.tex
\begin{abstract}

Recent advances in speech synthesis and voice conversion have made deepfake speech increasingly realistic, making generalization to unseen spoofing attacks a critical challenge. Pretrained speech and audio models offer a promising direction for improving robustness to such unseen attacks. Self-supervised learning (SSL) models capture fine-grained, low-level acoustic characteristics, whereas Auditory Large Language Models (ALLMs) provide higher-level contextual representations. These complementary views can provide useful cues for improving generalization to unseen attacks. However, direct fusion does not explicitly disentangle information shared across the two views from view-specific complementary information, limiting effective cross-view integration. To address this, we propose CRAF, a cross-view residual-aware fusion framework that uses ALLM-guided cross-view attention to enrich SSL representations and adopts ALLM as a high-level reference to separate ALLM-explainable information from complementary SSL residual information. The residual is selectively refined through adaptive gating and integrated through SSL-primary fusion. Experiments on ASVspoof 5 show that CRAF with Kimi-Audio achieves an EER of $5.96\%$ and a minDCF of $0.1192$, demonstrating robustness to unseen spoofing attacks.

\end{abstract}


%% file: sections/introduction_v2.tex
\section{Introduction}
Recent advances in speech synthesis and voice conversion have made deepfake speech increasingly realistic, posing growing risks to voice authentication and digital forensics
~\cite{kim2021conditional,todisco2019asvspoof}.
Deepfake detectors have evolved from handcrafted spectral and phase-based features~\cite{todisco2016new} to end-to-end architectures such as RawNet2~\cite{tak2021end} and AASIST~\cite{jung2022aasist}.
Despite strong performance under matched conditions, generalization to unseen synthesis methods, recording conditions, and datasets remains challenging~\cite{chen2020generalization,yamagishi2021asvspoof,pascu2024towards}.
This motivates representations that capture transferable spoofing characteristics rather than attack- or dataset-specific patterns.

\begin{figure*}[htbp]
\centering
\includegraphics[width=.95\linewidth]{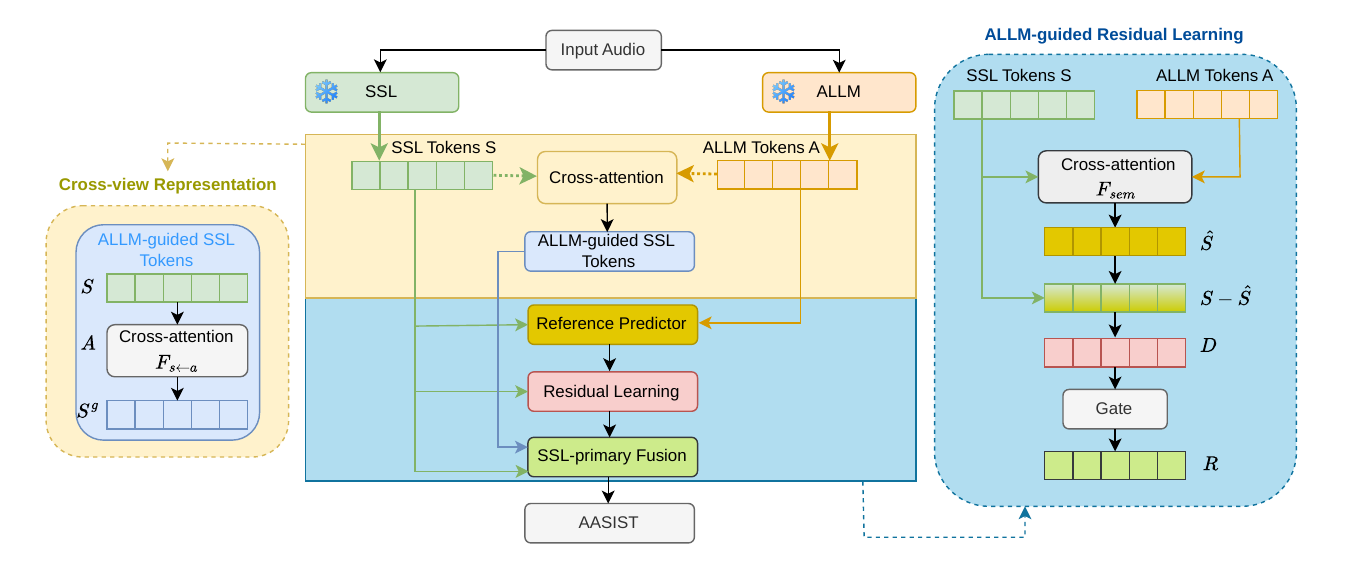}
\vspace{-20pt}
\caption{Overview of CRAF with ALLM-guided representation learning, cross-view residual learning, and SSL-primary fusion.}
\label{overall}
\end{figure*}

SSL models have become strong front-ends for deepfake speech detection because their learned representations provide informative features that are effective for distinguishing genuine and synthesized speech ~\cite{wang2025mixture,li2023voice}.
However, although SSL representations capture rich acoustic and contextual information, they are not explicitly designed to model the high-level semantic knowledge available from audio large language models.
Recent studies have therefore explored ALLMs for deepfake detection~\cite{gu2025allm4add,chuchra2025investigating}.
LLM representations provide rich semantic and contextual abstractions, but their semantic-dominant high-level representations overlook fine-grained acoustic cues important for spoof detection~\cite{guo2026towards}. Textual acoustic grounding further highlights this limitation by explicitly incorporating acoustic descriptors into LLM-based detectors to bridge the gap between subtle acoustic information and the LLM semantic space~\cite{kheir2026textual}.
Thus, SSL and ALLM provide complementary views: SSL preserves fine-grained acoustic characteristics, whereas ALLM provides higher-level contextual guidance.

The challenge is to exploit this complementarity without weakening acoustic information required for spoof detection.
A straightforward approach is to concatenate or directly fuse SSL and ALLM representations.
However, direct fusion treats both views uniformly and does not distinguish shared information from complementary SSL-specific characteristics~\cite{TAN2026132732}.
This can introduce redundancy while reducing emphasis on fine-grained SSL information weakly represented by the ALLM~\cite{LI2023612}.
Thus, the problem is not simply to combine SSL and ALLM representations, but to incorporate high-level guidance while preserving complementary SSL information.

To address this problem, we propose CRAF, a cross-view residual-aware fusion framework for deepfake speech detection. CRAF adopts an asymmetric design in which SSL serves as the primary representation and ALLM provides high-level guidance. First, ALLM-guided cross-view learning enriches the SSL representation with contextual information. CRAF then estimates SSL information recoverable under ALLM guidance and models the remaining SSL-specific information as a cross-view residual. The residual is adaptively refined to preserve complementary fine-grained characteristics. Finally, SSL-primary fusion selectively integrates guided and residual representations while retaining the original SSL stream.

The main contributions are threefold.
First, we introduce an asymmetric SSL--ALLM formulation that exploits high-level ALLM guidance while retaining SSL as the primary fine-grained representation.
Second, we propose ALLM-guided cross-view residual learning to capture complementary SSL information insufficiently represented by the ALLM view.
Third, we develop SSL-primary adaptive fusion to selectively integrate guided and residual information while preserving the original SSL representation.

%% file: sections/data.tex








%% file: sections/methodology_v4.tex
\section{Proposed Method}



The overall architecture of the proposed CRAF framework is illustrated in Fig.~\ref{overall}. The framework begins with frozen SSL and ALLM encoders that extract complementary fine-grained acoustic and high-level contextual representations, respectively. ALLM-guided cross-view attention then enriches the SSL stream, while an SSL reference estimator derives an ALLM-conditioned reference to isolate complementary SSL information. The resulting cross-view residual is selectively refined and combined with the guided SSL representation through SSL-primary adaptive fusion before AASIST-based classification.

\noindent\textbf{Cross-view Representation.}
Given an input waveform $x$, frozen SSL and ALLM encoders extract complementary
representations, which are projected into a shared $d$-dimensional latent space:
\begin{equation}
\begin{aligned}
    \mathbf{S} &= P_s\!\left(E_{\mathrm{SSL}}(x)\right), &
    \mathbf{A} &= P_a\!\left(E_{\mathrm{ALLM}}(x)\right),
\end{aligned}
\label{eq:encoder}
\end{equation}
where $E_{\mathrm{SSL}}(\cdot)$ and $E_{\mathrm{ALLM}}(\cdot)$ denote the
pretrained SSL and ALLM encoders, respectively, and $P_s(\cdot)$ and
$P_a(\cdot)$ are trainable projection layers. The projected representations are
$\mathbf{S}\in\mathbb{R}^{T_s\times d}$ and
$\mathbf{A}\in\mathbb{R}^{T_a\times d}$, where $T_s$ and $T_a$ denote their
sequence lengths. We set $d=256$, while both pretrained encoders remain frozen
throughout training.

To inject high-level ALLM information into the temporally fine-grained SSL
stream, we apply ALLM-guided cross-view attention:
\begin{equation}
    \mathbf{S}^{g}
    =
    F_{s\leftarrow a}(\mathbf{S},\mathbf{A}),
\label{eq:cross_attn}
\end{equation}
where $F_{s\leftarrow a}(\cdot)$ uses $\mathbf{S}$ as queries and
$\mathbf{A}$ as keys and values. The resulting
$\mathbf{S}^{g}\in\mathbb{R}^{T_s\times d}$ preserves the temporal resolution
of the SSL stream while incorporating contextual information from the ALLM
representation.

\noindent\textbf{ALLM-guided Residual Learning.}
Cross-view attention enriches the SSL stream with ALLM context, but does not
explicitly separate information already represented by the ALLM view from
complementary SSL information. CRAF addresses this by independently estimating
an ALLM-conditioned SSL reference and decomposing the SSL representation through
a cross-view residual:
\begin{equation}
\begin{aligned}
    \widehat{\mathbf{S}}
    &=
    F_{\mathrm{ref}}(\mathbf{S},\mathbf{A}), \\
    \mathbf{D}
    &=
    \mathrm{LN}
    \left(
        \mathbf{S}-\widehat{\mathbf{S}}
    \right),
\end{aligned}
\label{eq:reference_residual}
\end{equation}
where $F_{\mathrm{ref}}(\cdot)$ is an independently parameterized stack of
cross-attention layers using $\mathbf{S}$ as queries and $\mathbf{A}$ as keys
and values, and $\mathrm{LN}(\cdot)$ denotes layer normalization.
Although $F_{s\leftarrow a}$ and $F_{\mathrm{ref}}$ share the same
cross-attention formulation, they serve distinct functions:
$F_{s\leftarrow a}$ transfers contextual information into the SSL stream,
whereas $F_{\mathrm{ref}}$ estimates the portion of the SSL representation
explained by the ALLM view. Consequently,
$\mathbf{D}\in\mathbb{R}^{T_s\times d}$ isolates complementary SSL information
relative to the ALLM-conditioned reference.

To retain only detection-relevant residual information, CRAF applies a
feature-wise residual gate conditioned jointly on the original SSL
representation, the estimated reference, and their residual:
\begin{equation}
    \mathbf{R}
    =
    F_{\mathrm{enh}}
    \left(
        \sigma
        \left(
            F_r
            \left(
                [\mathbf{S},
                 \widehat{\mathbf{S}},
                 \mathbf{D}]
            \right)
        \right)
        \odot
        \mathbf{D}
    \right),
\label{eq:residual_learning}
\end{equation}
where $[\cdot]$ denotes feature-wise concatenation,
$F_r(\cdot)$ is a feed-forward gating network,
$\sigma(\cdot)$ is the sigmoid function,
$\odot$ denotes element-wise multiplication, and
$F_{\mathrm{enh}}(\cdot)$ is a feed-forward refinement network. The resulting
$\mathbf{R}\in\mathbb{R}^{T_s\times d}$ represents the selectively refined
cross-view residual.

\noindent\textbf{SSL-Primary Adaptive Fusion.}
CRAF combines the original SSL representation $\mathbf{S}$, the ALLM-guided SSL
representation $\mathbf{S}^{g}$, and the refined residual $\mathbf{R}$ while
explicitly preserving SSL as the primary stream. Feature-wise fusion weights
are computed as
\begin{equation}
    \mathbf{G}_{f}
    =
    \tanh
    \left(
        F_f
        \left(
            [\mathbf{S},
             \mathbf{S}^{g},
             \mathbf{R}]
        \right)
    \right),
\label{eq:fusion_gate}
\end{equation}
where $F_f(\cdot)$ denotes the fusion network. The final fused representation
is
\begin{equation}
    \mathbf{Z}
    =
    \mathbf{S}
    +
    \mathrm{Dropout}
    \left(
        \mathbf{G}_{f}
        \odot
        \left(
            \mathbf{R}
            +
            \lambda\mathbf{S}^{g}
        \right)
    \right),
\label{eq:adaptive_fusion}
\end{equation}
where $\lambda$ controls the contribution of the ALLM-guided SSL representation.
This formulation preserves the original SSL representation through a residual
path while adaptively injecting contextual information from
$\mathbf{S}^{g}$ and complementary cross-view information from $\mathbf{R}$.

\noindent\textbf{Training Objective.}
Given the fused representation $\mathbf{Z}$, the AASIST-based classifier
produces posterior probabilities
$\mathbf{p}_i=[p_{i,1},p_{i,2}]$ for bonafide and deepfake speech. The model is
optimized using weighted cross-entropy with label smoothing:
\begin{equation}
    \mathcal{L}_{\mathrm{cls}}
    =
    -\frac{1}{B}
    \sum_{i=1}^{B}
    \sum_{c=1}^{2}
    w_c\,\widetilde{y}_{i,c}\log p_{i,c},
\label{eq:loss}
\end{equation}
where $B$ is the mini-batch size, $w_c$ denotes the class-specific weight,
$\widetilde{y}_{i,c}$ is the label-smoothed target, and $p_{i,c}$ is the
predicted posterior probability for class $c$. The projection, cross-view
attention, reference estimation, residual learning, adaptive fusion, and
classification modules are jointly optimized, while the pretrained SSL and
ALLM encoders remain frozen.

%% file: sections/experiments.tex
\section{Experiments}

\begin{figure*}[ht]
    \centering
    \includegraphics[width=\linewidth,clip]{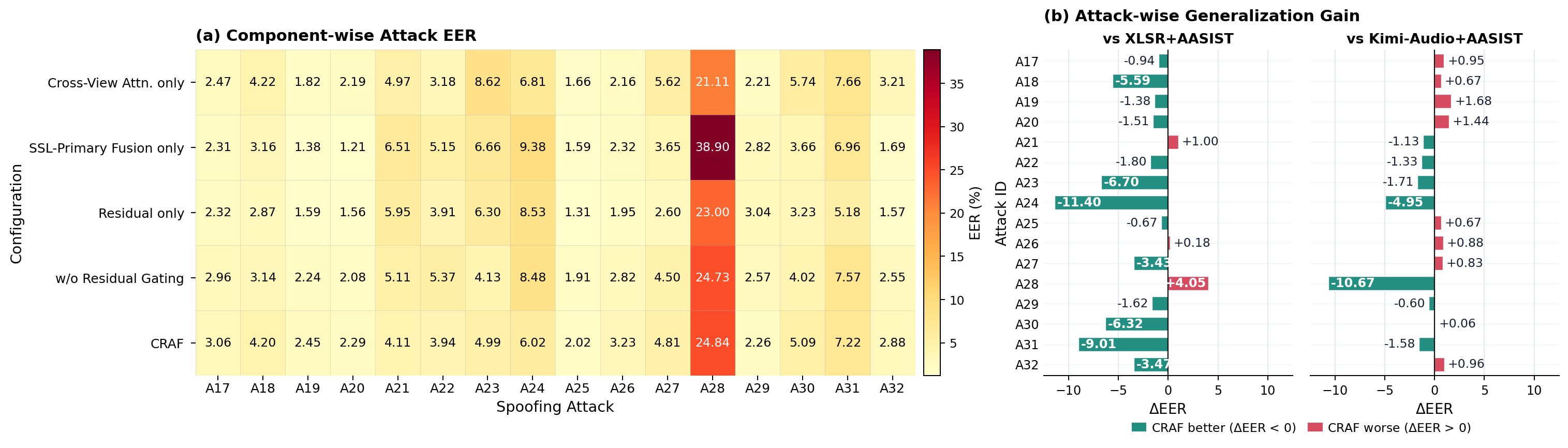}
    \vspace{-20pt}
    \caption{Attack-wise analysis on ASVspoof 5 eval. (a) EER (\%) across CRAF configurations. (b) $\Delta$EER of CRAF relative to XLS-R+AASIST and Kimi-Audio+AASIST.}
    \label{fig:embedding}
\end{figure*}



\subsection{Dataset and evaluation metics}

We used the ASVspoof $5$ dataset \cite{wang2024asvspoof} to evaluate the performance of the proposed method, as shown in Table~\ref{tb_asvspoof5}. We use the equal error rate (EER) and the minimum detection cost function (minDCF) as evaluation metrics.

\begin{table}[!t]
\centering
\caption{Number of utterances in the ASVspoof 5 dataset.}
\label{tb_asvspoof5}
\small
\begin{tabular}{llrrr}
\toprule
\multirow{2}{*}{\textbf{Dataset}} &
\multirow{2}{*}{\textbf{Splits}} &
\multicolumn{3}{c}{\textbf{Number of Utterances}} \\
& & \textbf{Genuine} & \textbf{Spoofed} & \textbf{Total} \\
\midrule
\multirow{3}{*}{ASVspoof 5}
& Training    & 18,797  & 163,560 & 182,357 \\
& Development & 31,334  & 109,616 & 140,950 \\
& Evaluation  & 138,688 & 542,086 & 680,774 \\
\bottomrule
\end{tabular}
\end{table}

\subsection{Experimental Setup}

The experimental setup includes audio preprocessing, frozen pretrained encoders,
CRAF modules, an AASIST-style backend, and training details. All audio is
resampled to $16\,\mathrm{kHz}$ and normalized. 
We use \texttt{XLS-R 300M}\footnotemark[1] as the SSL encoder and \texttt{Step-Audio}\footnotemark[2], \texttt{Qwen-Audio}\footnotemark[3], and \texttt{Kimi-Audio}\footnotemark[4] as the ALLM encoder. Designed for broad audio and speech understanding, these ALLMs provide higher-level semantic and contextual representations complementary to the acoustic representations captured by SSL models. The SSL and ALLM representations are projected into a shared $256$-dimensional space.
The cross-view attention and reference
estimator each use $2$ layers, $8$ heads, a feed-forward dimension of $1024$,
and $0.1$ dropout. Residual gating and SSL-primary fusion operate in the same
latent space. The AASIST-style backend uses $2$ graph layers with hidden
dimension $128$, pooling ratio $0.7$, and dropout $0.2$. Models are trained with
AdamW ($2\times10^{-5}$ learning rate, $10^{-3}$ weight decay, bfloat16) using
weighted cross-entropy with $0.05$ label smoothing, gradient accumulation of
$4$, and gradient clipping at $5.0$. The learning rate is halved on development
EER plateaus, and the best development checkpoint is used for evaluation.

\footnotetext[1]{%
\href{https://dl.fbaipublicfiles.com/fairseq/wav2vec/xlsr2_300m.pt}{XLS-R checkpoint},
\textsuperscript{2}\href{https://huggingface.co/stepfun-ai/Step-Audio-2-mini-Base}{Step-Audio-2-mini-Base},
\textsuperscript{3}\href{https://huggingface.co/Qwen/Qwen2-Audio-7B-Instruct}{Qwen2-Audio-7B-Instruct},
\textsuperscript{4}\href{https://huggingface.co/moonshotai/Kimi-Audio-7B-Instruct}{Kimi-Audio-7B-Instruct}.}

\subsection{Results}

\begin{table}[t]
\centering
\caption{Comparison with recent methods and our baselines on ASVspoof 5 Track 1 under the open condition.}
\label{tab:main_results}
\footnotesize
\setlength{\tabcolsep}{3.5pt}
\begin{tabular}{llccc}
\hline
\textbf{Category} & \textbf{Method} &
\multicolumn{2}{c}{\textbf{EER (\%)}} &
\textbf{minDCF} \\
\cline{3-4}
&& \textbf{Dev} & \textbf{Eval}& \textbf{Eval} \\
\hline

\multirow{6}{*}{\makecell{
\textit{Recent}\\
\textit{single-system}\\
\textit{methods}
}}
& ASTDT~\cite{wani2025astdt}                     & -- & 6.94  & -- \\
& ProSDD~\cite{mahapatra2026prosdd}              & -- & 7.38  & -- \\
& BiCrossMamba-ST~\cite{elkheir25_interspeech}   & -- & 29.53 & 0.6884
 \\
& Fused SSL + NeXt-TDNN~\cite{tahaoglu2025robust}
                                                  & -- & 7.23 & -- \\
& openSMILE~\cite{pascu2025easy}                 & -- & 15.70 & -- \\
 & C-MCSS-Mamba~\cite{app16178413}
                                                   & -- & 9.67  & -- \\
\hline

\multirow{4}{*}{\textit{Our baselines}}
& XLS-R + AASIST       & \textbf{1.22} & 9.78  & 0.1952 \\
& Step-Audio + AASIST  & 7.40 & 11.33 & 0.2223 \\
& Qwen-Audio + AASIST  & 6.76 & 12.37 & 0.2473 \\
& Kimi-Audio + AASIST  & 3.93 & 7.35  & 0.1437 \\
\hline

\multirow{3}{*}{\textit{Our proposed}}
& CRAF (Step-Audio)    & 4.65 & 8.45 & 0.1679 \\
& CRAF (Qwen-Audio)    & 4.20 & 8.07 & 0.2033 \\
& CRAF (Kimi-Audio)    & 2.41 & \textbf{5.96} & \textbf{0.1192} \\
\hline

\end{tabular}
\end{table}

The results in Table~\ref{tab:main_results} demonstrate the effectiveness of CRAF across different ALLM encoders. All CRAF variants improve over their corresponding single-encoder baselines on the evaluation set, indicating that combining fine-grained SSL representations with higher-level ALLM information provides complementary benefits for deepfake speech detection. Among the evaluated configurations, CRAF (Kimi-Audio) achieves the best performance, with an evaluation EER of $5.96\%$ and a minDCF of $0.1192$, improving over both the Kimi-Audio and XLS-R baselines.

Although XLS-R + AASIST obtains the lowest development EER, its performance degrades considerably on the evaluation set. In contrast, CRAF shows stronger evaluation performance, suggesting improved robustness to unseen spoofing conditions. Moreover, the consistent gains observed with Step-Audio, Qwen-Audio, and Kimi-Audio indicate that the effectiveness of CRAF is not dependent on a specific ALLM encoder, but instead results from the proposed cross-view interaction and residual-aware fusion.

CRAF (Kimi-Audio) also achieves the best evaluation EER among the recent single-system methods considered in this comparison. Notably, CRAF is evaluated as a single model without model- or score-level ensembling. Overall, these results support the effectiveness of CRAF for robust deepfake speech detection under unseen conditions.

\subsection{Ablation Studies}

\begin{table}[t]
\centering
\caption{Ablation study of CRAF components.}
\label{tab:ablation}
\footnotesize
\setlength{\tabcolsep}{4pt}
\begin{tabular}{llccc}
\hline
\textbf{Category} & \textbf{Method} &
\multicolumn{2}{c}{\textbf{EER (\%)}} &
\textbf{minDCF} \\
\cline{3-4}
&& \textbf{Dev} & \textbf{Eval}& \textbf{Eval} \\
\hline
XLS-R       & AASIST                    & \textbf{1.22} & 9.78  & 0.1952 \\
Step-Audio  & AASIST                    & 7.40 & 11.33 & 0.2223 \\
Qwen-Audio  & AASIST                    & 6.76 & 12.37 & 0.2473 \\
Kimi-Audio  & AASIST                    & 3.93 & 7.35  & 0.1437 \\
\hline
Step-Audio  & SSL-Primary Fusion only   & 6.58 & 8.68 & 0.1675 \\
            & Cross-view Attention only & 2.87 & 7.98 & 0.1593 \\
            & Residual only             & 6.26 & 9.41 & 0.1876 \\
\hline
Qwen-Audio  & SSL-Primary Fusion only   & 4.29 & 8.35 & 0.1666 \\
            & Cross-view Attention only & 3.04 & 8.41 & 0.1678 \\
            & Residual only             & 4.97 & 8.59 & 0.1708 \\
\hline
Kimi-Audio  & SSL-Primary Fusion only   & 2.70 & 7.12 & 0.1411 \\
            & Cross-view Attention only & 1.88 & 6.71 & 0.1342 \\
            & Residual only             & 1.99 & 6.65 & 0.1328 \\
            & w/o Residual Gating       & 3.01 & 6.69 & 0.1335 \\
\hline
Kimi-Audio  & CRAF                      & 2.41 & \textbf{5.96} & \textbf{0.1192} \\
\hline
\end{tabular}
\end{table}

The ablation results in Table~\ref{tab:ablation} show that cross-view attention, residual learning, and SSL-primary fusion each contribute to the effectiveness of CRAF across different ALLM encoders. With Kimi-Audio, the full CRAF achieves the best performance, reaching an EER of $5.96\%$ and a minDCF of $0.1192$, outperforming all individual component variants. The consistent degradation observed when using only one component indicates that these modules provide complementary benefits rather than redundant functionality. Removing residual gating also increases the EER to $6.69\%$ and the minDCF to $0.1335$, confirming the importance of selectively refining residual information. Overall, the consistent trends in both EER and minDCF demonstrate that jointly integrating the three components improves robustness to unseen conditions.

\subsection{Attack-wise Analysis}


The individual CRAF configurations exhibit distinct behaviors across spoofing attacks, as shown in Fig.~\ref{fig:embedding}(a). No single component is consistently optimal across all attacks: the full CRAF achieves the lowest EER among the compared configurations on A21 and A24, while individual components perform better on several other attacks. A28 remains particularly challenging, with EERs ranging from 21.11\% to 38.90\%. These variations show that ALLM-guided cross-view attention, residual learning, and SSL-primary fusion respond differently to individual spoofing attacks, suggesting that each component contributes distinct attack-dependent information and plays a
complementary role within the full framework.


CRAF also exhibits stronger attack-wise generalization than the single-view baselines, as illustrated in Fig.~\ref{fig:embedding}(b). Compared with XLS-R+AASIST, CRAF reduces EER on 13 of 16 attacks, with the largest improvements on A24 ($-11.40$), A31 ($-9.01$), A23 ($-6.70$), A30 ($-6.32$), and A18 ($-5.59$) percentage points. Compared with Kimi-Audio+AASIST, the gains are more attack-dependent, with improvements on 7 of 16 attacks, most notably A28 ($-10.67$) and A24 ($-4.95$). These results show that CRAF provides substantial gains over the SSL-only baseline while improving the stronger ALLM-only baseline on specific attacks, demonstrating that jointly exploiting both representation views improves robustness across diverse unseen spoofing attacks.

%% file: sections/conclusion.tex
\section{Conclusion}


This work proposed CRAF, a cross-view residual-aware fusion framework for deepfake speech detection that jointly exploits complementary SSL and ALLM representations. By using the ALLM representation as a high-level reference, CRAF preserves complementary SSL residual information and adaptively integrates it with ALLM-guided SSL features. Experiments on ASVspoof 5 showed improved performance over the individual encoders and direct fusion baseline, demonstrating the effectiveness of CRAF for robust detection of diverse and unseen spoofing attacks. Future work will investigate improved cross-view interaction and generalization across different deepfake datasets.

%% file: refs.bib
@inproceedings{kim2021conditional,
  title={Conditional variational autoencoder with adversarial learning for end-to-end text-to-speech},
  author={Kim, Jaehyeon and Kong, Jungil and Son, Juhee},
  booktitle={Proc. ICML},
  pages={5530--5540},
  year={2021}
}

@inproceedings{todisco2019asvspoof,
  title={ASVspoof 2019: Future Horizons in Spoofed and Fake Audio Detection},
  author={Todisco, Massimiliano and Wang, Xin and Vestman, Ville and Sahidullah, Md and Delgado, H{\'e}ctor and Nautsch, Andreas and Yamagishi, Junichi and Evans, Nicholas and Kinnunen, Tomi and Lee, Kong Aik},
  booktitle={Interspeech 2019},
  year={2019}
}

@inproceedings{wang2025mixture,
  title={Mixture of experts fusion for fake audio detection using frozen wav2vec 2.0},
  author={Wang, Zhiyong and Fu, Ruibo and Wen, Zhengqi and Tao, Jianhua and Wang, Xiaopeng and Xie, Yuankun and Qi, Xin and Shi, Shuchen and Lu, Yi and Liu, Yukun and others},
  booktitle={Proc. ICASSP},
  pages={1--5},
  year={2025}
}

@article{li2023voice,
  title={Voice deepfake detection using the self-supervised pre-training model hubert},
  author={Li, Lanting and Lu, Tianliang and Ma, Xingbang and Yuan, Mengjiao and Wan, Da},
  journal={Applied Sciences},
  volume={13},
  number={14},
  pages={8488},
  year={2023}
}

@inproceedings{jung2022aasist,
  title={Aasist: Audio anti-spoofing using integrated spectro-temporal graph attention networks},
  author={Jung, Jee-weon and Heo, Hee-Soo and Tak, Hemlata and Shim, Hye-jin and Chung, Joon Son and Lee, Bong-Jin and Yu, Ha-Jin and Evans, Nicholas},
  booktitle={Proc. ICASSP},
  pages={6367--6371},
  year={2022},
}

@inproceedings{gu2025allm4add,
  title={Allm4add: Unlocking the capabilities of audio large language models for audio deepfake detection},
  author={Gu, Hao and Yi, Jiangyan and Wang, Chenglong and Tao, Jianhua and Lian, Zheng and He, Jiayi and Ren, Yong and Chen, Yujie and Wen, Zhengqi},
  booktitle={Proc. ACMMM},
  pages={11736--11745},
  year={2025}
}

@inproceedings{chuchra2025investigating,
  title={Investigating the Viability of Employing Multi-modal Large Language Models in the Context of Audio Deepfake Detection},
  author={Chuchra, Akanksha and Reddy, Shukesh and Mishra, Sudeepta and Das, Abhijit and Dhall, Abhinav},
  booktitle={Proc. IJCB},
  pages={1--11},
  year={2025}
}

@inproceedings{wang2024asvspoof,
  title={ASVspoof 5: crowdsourced speech data, deepfakes, and adversarial attacks at scale},
  author={Wang, Xin and Delgado, H{\'e}ctor and Tak, Hemlata and Jung, Jee-weon and Shim, Hye-jin and Todisco, Massimiliano and Kukanov, Ivan and Liu, Xuechen and Sahidullah, Md and Kinnunen, Tomi H and others},
  booktitle={Proc. ASVspoof 2024},
  pages={1--8},
  year={2024}
}

@inproceedings{chen2020generalization,
  title={Generalization of audio deepfake detection},
  author={Chen, Tianxiang and Kumar, Avrosh and Nagarsheth, Parav and Sivaraman, Ganesh and Khoury, Elie},
  booktitle={Proc. Odyssey 2020},
  pages={132--137},
  year={2020}
}

@inproceedings{yamagishi2021asvspoof,
  title={ASVspoof 2021: Accelerating progress in spoofed and deepfake speech detection},
  author={Yamagishi, Junichi and Wang, Xin and Todisco, Massimiliano and Sahidullah, Md and Patino, Jose and Nautsch, Andreas and Liu, Xuechen and Lee, Kong Aik and Kinnunen, Tomi and Evans, Nicholas and Delgado, H{\'e}ctor},
  booktitle={Proc. ASVspoof},
  pages={47--54},
  year={2021},
  doi={10.21437/ASVSPOOF.2021-8}
}

@inproceedings{todisco2016new,
  title={A new feature for automatic speaker verification anti-spoofing: Constant Q cepstral coefficients},
  author={Todisco, Massimiliano and Delgado, H{\'e}ctor and Evans, Nicholas},
  booktitle={Proc. Odyssey 2016},
  pages={283--290},
  year={2016}
}

@inproceedings{tak2021end,
  title={End-to-end anti-spoofing with rawnet2},
  author={Tak, Hemlata and Patino, Jose and Todisco, Massimiliano and Nautsch, Andreas and Evans, Nicholas and Larcher, Anthony},
  booktitle={Proc. ICASSP},
  pages={6369--6373},
  year={2021}
}

@inproceedings{pascu2024towards,
  title={Towards generalisable and calibrated audio deepfake detection with self-supervised representations},
  author={Pascu, Octavian and Stan, Adriana and Oneata, Dan and Oneata, Elisabeta and Cucu, Horia},
  booktitle={Proc. Interspeech},
  pages={4828--4832},
  year={2024}
}

@inproceedings{mahapatra2026prosdd,
  author    = {A. Mahapatra and I. R. Ulgen and K. A. Lee and N. Andrews and B. Sisman},
  title     = {ProSDD: Learning Prosodic Representations for Speech Deepfake Detection Against Expressive and Emotional Attacks},
  booktitle = {Proc. Interspeech},
  year      = {2026}
}

@article{wani2025astdt,
  title={ASTDT: an interpretable adaptive spectro-temporal diffusion transformer for audio deepfake detection},
  author={Wani, Taiba Maijd and Qadri, Syed Asif Ahmad and Ashraf, Arselan and Amerini, Irene},
  journal={EURASIP Journal on Information Security},
  volume={2025},
  number={1},
  pages={32},
  year={2025},
}

@inproceedings{elkheir25_interspeech,
  title     = {{BiCrossMamba-ST: Speech Deepfake Detection with Bidirectional Mamba Spectro-Temporal Cross-Attention}},
  author    = {Yassine {El Kheir} and Tim Polzehl and Sebastian Möller},
  year      = {2025},
  booktitle = {Proc. Interspeech},
  pages     = {2235--2239},
}

@article{tahaoglu2025robust,
  title={Robust DeepFake audio detection via an improved NeXt-TDNN with multi-fused self-supervised learning features},
  author={Tahaoglu, Gul},
  journal={Applied Sciences},
  volume={15},
  number={17},
  pages={9685},
  year={2025}
}

@inproceedings{pascu2025easy,
  title={Easy, Interpretable, Effective: openSMILE for voice deepfake detection},
  author={Pascu, Octavian and Onea{\c{t}}{\u{a}}, Dan and Cucu, Horia and M{\"u}ller, Nicolas},
  booktitle={Proc. ICASSP},
  pages={1--5},
  year={2025},
}

@article{LI2023612,
title = {Instance-wise multi-view representation learning},
journal = {Information Fusion},
volume = {91},
pages = {612-622},
year = {2023},
issn = {1566-2535},
doi = {https://doi.org/10.1016/j.inffus.2022.11.006},
author = {Dan Li and Haibao Wang and Yufeng Wang and Shengpei Wang},
}

@Article{app16178413,
AUTHOR = {Zhang, Gaopeng and Liu, Shidong and Zhang, Dengtao and Tang, Liang},
TITLE = {C-MCSS-Mamba: Counterfactual Mechanism-Contrastive Selective Scan Within Mamba for Block-Causal Speech Deepfake Detection},
JOURNAL = {Applied Sciences},
VOLUME = {16},
YEAR = {2026},
NUMBER = {17},
ARTICLE-NUMBER = {8413},
ISSN = {2076-3417},
DOI = {10.3390/app16178413}
}

@article{kheir2026textual,
  title={Textual Acoustic Grounding for Generalizable LLM-Based Deepfake Voice Detection},
  author={Kheir, Yassine El and Wang, Xin and Ge, Wanqing and Polzehl, Tim and Moeller, Sebastian and Yamagishi, Junichi},
  journal={arXiv preprint arXiv:2608.30622},
  year={2026}
}

@article{TAN2026132732,
title = {Mutual-guidance framework for audio DeepFake detection via multi-dimensional feature interaction},
journal = {Neurocomputing},
volume = {672},
pages = {132732},
year = {2026},
author = {Dengtai Tan and Boao Tan and Deyi Yang and Yang Yang and Chengyu Niu},
}

@article{guo2026towards,
  title={Towards Explicit Acoustic Evidence Perception in Audio LLMs for Speech Deepfake Detection},
  author={Guo, Xiaoxuan and Xie, Yuankun and Cheng, Haonan and Zhou, Jiayi and Liu, Jian and Huang, Hengyan and Ye, Long and Zhang, Qin},
  journal={arXiv preprint arXiv:2601.23066},
  year={2026}
}
